\documentclass[12pt,english]{article}
\usepackage[T1]{fontenc}
\usepackage[latin9]{inputenc}
\usepackage{geometry}
\usepackage{amsmath}
\usepackage{amssymb}
\usepackage{setspace}

\makeatletter
\newcommand{\lyxaddress}[1]{
\par {\raggedright #1
\vspace{1.4em}
\noindent\par}
}

\pdfoutput=1

\usepackage{babel}
\date{ }

\makeatother

\usepackage{babel}
\begin{document}

\title{\noindent A relational approach to the Mach-Einstein question}

\author{{\normalsize{Herman Telkamp}}%
\thanks{Jan van Beverwijckstraat 104, 5017 JA, The Netherlands, email: herman\_telkamp@hotmail.com%
}}

\maketitle

\lyxaddress{\begin{center}
{\small{Essay written for the Gravity Research Foundation 2013 Awards}}\\
{\small{for Essays on Gravitation}}%
\footnote{{\small{Received honorable mention}}%
}{\small{}}\\
{\small{\bigskip{}
}}{\footnotesize{Submitted March 5, 2013}}
\par\end{center}}
\begin{abstract}
\noindent Mach's principle is incompatible with general relativity
(GR), it has not condensed into an established theory and suffers
from inconsistencies. Yet, the problem is that Mach's principle is
a consequence of Berkeley's notions, which are as good as irrefutable
for their ontological nature. Moreover, the observed coincidence of
the \textquotedbl{}preferred inertial frame\textquotedbl{} and the
frame attached to the \textquotedbl{}fixed stars\textquotedbl{} is
essentially Machian, while this coincidence is anomalous to both GR
and Newtonian physics. Another issue is that GR needs dark energy
to explain the accelerating expansion of the universe, while acceleration
of receding masses is inherent to the Machian principle. So GR and
Mach's principle question each other, while neither one can be falsified
easily. This suggest that both are valid in their particular domain.
A relational theory may reconcile the two, since it can cover both.
\end{abstract}
\medskip{}

\noindent An example of a physical \textit{unobservable} is the velocity
of a single body in otherwise empty space. The unobservable is not
involved in any physical relationship, therefore is physically meaningless,
or inexistent, for that matter. Entities such as distance, velocity
and time arise as relational properties only from the presence of
a second body. This notion is fundamental to relational physics, advocating
that physical properties can only emerge from the interaction of objects.
This \textquotedbl{}relational principle\textquotedbl{} includes Mach's
principle (inertia emerges from the distribution of matter), but is
much wider. Also space and time need matter to exist. The relational
principle implies that physical properties disappear as an object
gets isolated from anything else. As obvious as this is for the force
of gravity, it is not obvious at all for inertia, space and time,
which we deem absolute (Newton) or at best curvable by mass-energy
(Einstein). GR clearly embodies aspects of the relational principle,
as the spacetime metric depends on mass distribution, and length contraction
can be interpreted as increase of inertia. Even so, in GR these properties
do not vanish, but remain absolute in empty space. This must have
been one reason why Einstein was unsuccessful in his attempts to incorporate
Mach's Principle in GR; the two are incompatible in empty space. This
was already noted by De Sitter back in 1917 \cite{Sitter (1917)}
and finally (after a long debate) acknowledged by Einstein. 

GR is a well founded theory, practically raised above any suspicion,
while Mach's principle has never really escaped the stage of concept
and suffers from inconsistencies. Yet, Mach's principle is consistent
with Berkeley's notions on gravity, while GR is not. The problem is
that Berkeley's notions (discussed hereafter) are as good as irrefutable
for their logic, simplicity and ontological nature, which I believe
has not been fully appreciated by the community. And so is Mach's
principle, being a consequence of Berkeley's logic. Apart from its
ontological foundation, Mach's principle is also supported empirically;
relativistic trajectories, like the perihelion precession and frame-dragging
can be derived from it in a straightforward manner \cite{Schroedinger (1926),Telkamp (2012)}.
Moreover, the preferred inertial frame for the solar system coincides
to high accuracy with the ICRF, the frame linked to extremely distant
radio sources, constituting the modern interpretation of the \textquotedbl{}fixed
stars\textquotedbl{} \cite{Kovalevsky (1997)}. While this is essentially
Machian, it is in fact an anomalous coincidence within the scope of
both Newtonian theory and GR, even though probably nobody doubted
that absolute space or Minkowski spacetime (notably GR's empty space
solution) are somehow connected to the fixed stars. But it just doesn't
follow from these theories. 

Thus, Mach's principle, as useless as it is, makes a case. This leaves
us in discomfort, as GR and Mach's principle question each other,
but neither one can be falsified easily. This suggests that both are
valid, but in different (yet overlapping) domains. It is necessary
to revive the debate and reconcile these conflicting treasures of
science. This essay attempts to bring some motion in this inert matter.
First, I introduce a realization of Mach's principle, following Schrödinger's
approach \cite{Schroedinger (1926)}, and point at its abilities and
limitations. Next, we will consider both GR and Mach's principle from
the perspective of the relational principle, which has the capacity
to cover both. I will argue that (only) a relational theory provides
a resolution and that Berkeley's notions guide the way.

\subsection*{Machian physics}

Of particular interest is Berkeley's essay \textit{De Motu} \cite{Berkeley (1721)}
on matter, space and time%
\footnote{Berkeley's essay was submitted unsuccessfully for a prize of the French
\textit{Académie des Sciences}%
}. His criticism of Newton's concept of absolute space regards the
notion that motion of a single object (point mass $m_{1}$) in empty
space is unobservable. If velocity is physically inexistent in the
one-body universe, then the same applies to the kinetic energy of
the single object. So, how could this object exhibit mass inertia?
In agreement with Mach's principle, it can not. This changes when
a second point mass $m_{2}$ appears. Due to gravity, the two bodies
will accelerate toward each other and build up kinetic energy. Therefore
$m_{1}$ and $m_{2}$ \textit{must} have acquired some inertia due
to each others presence; thus Mach's principle follows from Berkeley's
logic. Moreover, and this is key, the emergent inertia is manifest
in the \textit{radial} motion only. In Berkeley's view, contrary to
Newton's, the radial distance is the only meaningful geometrical parameter
among the two bodies. Indeed, as pointed out by Berkeley, any circular
motion of the two bodies around each other in empty space is physically
meaningless; our frame of reference could as well be rotating in the
opposite direction. Therefore, by this logic, which we will label
the \textit{anisotropic principle} of Berkeley, \textit{only} radial
motion matters in the relationship of the two bodies. Ergo, motion
perpendicular to the radial direction has no inertia and does not
represent kinetic energy between the two bodies. 

\medskip{}

\noindent From Berkeley's anisotropic principle we can draw some very
interesting conclusions: the Newtonian kinetic energy attributed to
the circular orbit of two bodies \textit{in empty space} is all virtual,
because unobservable. Thus, the 'real' (Machian) kinetic energy of
the revolving system ($m_{1}$,$m_{2}$), denoted $T_{12}$, is zero.
If, however, $m_{1}$ and $m_{2}$ were in a non-circular orbit, this
would involve radial motion between these bodies, therefore $T_{12}>0$.
Machian kinetic energy may be interpreted as the part of the Newtonian
kinetic energy that would be dissipated in an inelastic collision.
Indeed, \textquotedbl{}freezing\textquotedbl{} the two objects together
stops any relative radial motion, but it does not affect the rotation
or translation of the total system in empty space. These latter motions
are unobservable, therefore this part of Newtonian kinetic energy
is virtual. So, how can we reconcile this with Newtonian physics?

The above picture changes if we move $m_{1}$ and $m_{2}$ from empty
space into our universe, which we may represent by a hollow sphere
of mass $m_{o}$. Then, the same circular orbit of $m_{1}$ and $m_{2}$,
implies (components of) radial motion of both $m_{1}$ and $m_{2}$
relative to the masses which together constitute the surrounding spherical
shell $m_{o}$. So the orbit of the two bodies involves kinetic energies
$T_{01}$ and $T_{02}$ of the subsystems ($m_{1}$,$m_{o}$) and
($m_{2}$,$m_{o}$), respectively. As Schrödinger shows \cite{Schroedinger (1926)},
these two terms actually represent the Newtonian kinetic energies
of $m_{1}$ and $m_{2}$, provided velocities are relative to a frame
attached to $m{}_{o}$, the \textquotedbl{}fixed stars\textquotedbl{}. 

\medskip{}

\noindent Next to these Newtonian terms we still have $T_{12}$ as
a (very small) extra energy term arising from the interaction of the
two local bodies. This term is entirely responsible for relativistic
trajectories, like the perihelion precession \cite{Schroedinger (1926),Telkamp (2012)}
and frame-dragging \cite{Telkamp (2012)}. These results are based
on the following frame independent definition of the \textit{Machian
kinetic energy} of point masses $m_{i}$ and $m_{j}$ 
\begin{equation}
T_{ij}={\scriptstyle \frac{1}{2}}\mu_{ij}\dot{r}_{ij}^{2},\label{eq:Tij-1}
\end{equation}
where $r_{ij}$ denotes their separation and where $\mu_{ij}$ is
the \textit{partial inertia} between the masses $m_{i}$ and $m_{j}$
\begin{equation}
\mu_{ij}=\frac{-Gm_{i}m_{j}}{\varphi_{o}r_{ij}}=m_{i}\frac{\varphi_{j}(r_{i})}{\varphi_{o}}=m_{j}\frac{\varphi_{i}(r_{j})}{\varphi_{o}}.\label{eq:muij-1}
\end{equation}
$G$ is Newton's constant and $\varphi_{o}$ is a constant scaling
factor, equal to the gravitational background potential of the universe.
$\varphi_{j}(r_{i})$ denotes the potential due to $m_{j}$ at the
position of $m_{i}$. The definition of inertia (\ref{eq:muij-1})
satisfies the relational principle; it vanishes at infinite separation
and both inertia and kinetic energy are defined as mutual, frame independent
properties between each pair of bodies, just like the force of gravity
and potential energy. Classical gravitational potential energy between
the two bodies can be expressed as
\begin{equation}
V_{ij}=\frac{-Gm_{i}m_{j}}{r_{ij}}=\mu_{ij}\varphi_{o}.\label{eq:Vij}
\end{equation}
Hence, inertia\textit{ is} potential energy, thus giving interpretation
to the mass-energy equi\-valence. Furthermore, the correct prediction
of relativistic trajectories consistently requires \cite{Telkamp (2012)}
\begin{equation}
\varphi_{o}={\scriptstyle -\frac{1}{2}}c^{2}.\label{eq:phio}
\end{equation}
The total energy of an isolated system of point masses is straightforwardly
the sum over all pairs
\begin{equation}
E\;=\;\underset{{\scriptstyle {\scriptscriptstyle i}}}{{\textstyle \sum}}\underset{{\scriptstyle {\scriptscriptstyle j>i}}}{{\textstyle \sum}}\;\; T_{ij}+V_{ij}\;=\underset{{\scriptstyle {\scriptscriptstyle i}}}{\;{\textstyle \sum}}\underset{{\scriptstyle {\scriptscriptstyle j>i}}}{{\textstyle \sum}}\;\;{\scriptstyle \frac{1}{2}}\mu_{ij}\dot{r}_{ij}^{2}+\mu_{ij}\varphi_{o}\;=\underset{{\scriptstyle {\scriptscriptstyle i}}}{\;{\textstyle \sum}}\underset{{\scriptstyle {\scriptscriptstyle j>i}}}{{\textstyle \sum}}\;\;{\scriptstyle \frac{1}{2}}\mu_{ij}(\dot{r}_{ij}^{2}-c^{2}).\label{eq:E-1}
\end{equation}
(Note that the right hand side reflects the inherent relativistic
properties of Machian inertia \cite{Telkamp (2012)}). Since every
object can be considered composed of infinitesimal small point masses,
the above definitions are generic, i.e. hold for any mass distribution.
One can conveniently derive compound expressions for any two finite
size objects, translating or spinning relative to each other \cite{Telkamp (2012),Schroedinger (1926)}.
Notably, one obtains for a mass $m_{i}$ moving inside of the cosmic
hollow sphere $m_{o}$ with flat internal potential $\varphi_{b}=\varphi_{o}$,

\vspace{0.25cm}

\begin{spacing}{0.5}
\noindent 
\begin{equation}
\mu_{oi}=m_{i}\frac{\varphi_{b}}{\varphi_{o}}=m_{i}\label{eq:muio}
\end{equation}

\end{spacing}

\begin{spacing}{0.80000000000000004}
\noindent and
\end{spacing}

\noindent 
\begin{equation}
T_{oi}={\scriptstyle \frac{1}{2}}m_{i}v_{oi}^{2}.\label{eq:Tnewton}
\end{equation}
where $v_{oi}$ is the velocity of $m_{i}$ in the frame attached
to $m_{o}$. Thus, Newtonian physics follows directly from the anisotropic
Machian relations, \textit{provided} that one assumes the cosmic masses
present. In other words: the background potential $\varphi_{o}$ is
implicitly assumed in Newtonian physics via the fixed value of inertia
$\mu_{oi}=m_{i}$ of each body relative to the cosmos. 

The same must be true for GR, as Newtonian mechanics are recovered
in GR's weak-field limit. As a result, the implicit cosmic potential
prevents inertia (and spacetime alike) to vanish at \textquotedbl{}vacuum\textquotedbl{}
infinity. Hence, at non-relativistic speeds, absolute Minkowski spacetime
coincides with Newton's absolute space, which is the Machian space
of the large scale universe, homogeneously filled with matter. This
implies a restriction to GR: it (only) holds wherever the background
potential equals $\varphi_{o}$. Fortunately, we live in such a universe
and one can argue that in any other spacetime the same speed of light,
and so the same potential $\varphi_{b}=\varphi_{o}={\scriptstyle -\frac{1}{2}}c^{2}$,
will be measured, locally. Therefore, notwithstanding (\ref{eq:muio}),
the equivalence principle will hold locally.

A problem arises in cosmology, though, as it can not be treated locally.
The cosmic potential declines with the expansion of the universe.
Assuming gravitational interaction propagates at finite speed, (\ref{eq:muio})
implies that the inertia of matter farther away is higher due to a
higher local background potential at earlier epochs. From a Machian
point of view, this must somehow be accounted for, which however doesn't
seem to be the case in FRW cosmology. Or, is this problem called dark
energy? This suggestion is perhaps not as blunt as it appears; acceleration
of receding masses is inherent to the Machian principle: declining
background potential $\varphi_{b}$ results in decreasing inertia
(\ref{eq:muio}), therefore in acceleration of the cosmic expansion
\cite{Telkamp (2012)}. In GR, though, the metric is constrained,
i.e. does not vanish due to the implicit fixed background potential.
So something is needed (dark energy) to counteract this ever present
potential $\varphi_{o}$.

\medskip{}

\noindent Despite the beautiful consistency of the above framework,
Mach's principle has always suffered from an inconsistency problem.
Anisotropic inertia satisfies Berkeley's principle and anisotropy
is indispensable in predicting relativistic trajectories. However,
it also gives rise to anisotropic time dilation (the Machian harmonic
oscillator cycles slower in the direction of a mass kernel), which
is inconsistent with GR's isotropic time dilation. Moreover, as we
know since the famous Hughes-Drever experiments \cite{Drever (1961)},
clocks indeed appear not sensitive to direction. To many, this presented
conclusive proof of the isotropy of inertia. This is essentially where
the Machian doctrine got stuck, while GR passed all the tests.

\medskip{}

\noindent So a Machian theory has to deal with seemingly contradicting
requirements. It must be isotropic and anisotropic at the same time,
while anisotropy (i.e. Berkeley's principle) is considered conflicting
with Hughes-Drever experiments. One must realize, however, that no
experiment can invalidate Berkeley's principle, as it stands above
any experiment due to its ontological nature. Therefore, instead of
dismissing Mach's principle, we should try harder fixing the theory.
In view of the above, a resolution would involve a relational theory
comprising of both an anisotropic and an isotropic component, just
like in GR, but without the limiting implicit assumption of a fixed
background potential. The anisotropic (Machian) component serves relativistic
trajectories, while the isotropic part covers effects of remote observation,
like time dilation. An ansatz follows.

\subsection*{Relational physics}

\noindent Emergence of space and time must somehow happen at the point
where these entities become observable, so at the point where the
second body appears in empty space. This means spacetime can not actually
exist in an empty universe, nor in the universe of a single point
mass%
\footnote{A finite size body would represent multiple point masses, i.e. is
not precisely a single mass.%
}. By Berkeley's principle one can argue that a two-body universe has
only one spatial dimension, the line connecting the two point masses,
plus the dimension of time. The other two spatial dimensions arise
from the appearance of bodies in other directions. An ensemble of
four bodies can thus form a 3+1 dimensional spacetime. However, if
we step back and look from an increasing distance, then this collection
of bodies would gradually shrink into a single point mass in an empty
space, making space and time gradually dilute at larger scales and
ultimately vanish at infinity. Thus, asymptotically, an arbitrary
mass distribution in a finite volume can not be distinguished from
a single point mass, meaning spacetime has gone at vacuum infinity!
So we must contemplate a mechanism of transition; at a large but finite
distance from all mass, space and time must have partially lost their
significance. 

This regards the question how a relational metric of spacetime actually
looks like. There must be a clue in GR, since it correctly describes
relativistic phenomena wherever the background potential is $\varphi_{o}$.
But it does so by varying unit length and unit time only, while in
a relational sense we expect also inertia to vary. There is also a
clue in the Machian approach, where inertia varies along with potential,
but which lacks a spacetime metric. So we may obtain this missing
relational metric by analyzing the difference between a (correct)
GR solution and the (incomplete) Machian equation for the same case.
We compare the energy equation associated with the Schwarzschild metric
and the Machian energy equation for the same configuration: a small
test particle $m$ orbiting a massive sphere $M$ against a background
potential $\varphi_{o}$. Using definitions (\ref{eq:Tij-1})..(\ref{eq:Tnewton}),
the Machian energy equation in proper polar coordinates ($\rho,\phi$)
follows,
\begin{equation}
T_{mM}+T_{mo}+V_{mM}+V_{mo}={\scriptstyle \frac{1}{2}}\mu\overset{\,{\scriptscriptstyle \circ}}{\rho}\phantom{\hphantom{}}^{2}+{\scriptstyle \frac{1}{2}}m(\overset{\,{\scriptscriptstyle \circ}}{\rho}\phantom{\hphantom{}}^{2}+\rho^{2}\overset{\:{\scriptscriptstyle \circ}}{\phi}\phantom{\hphantom{}}^{2})+\mu\varphi_{o}+m\varphi_{o}=E,\label{eq:Mach energy}
\end{equation}
where $\mu=\mu_{mM}=-GmM/\varphi_{o}\rho$ is the partial inertia
between the orbiting masses, while $m$ represents the partial inertia
between the particle and the cosmic background. The circle $^{{\scriptstyle \circ}}$
denotes the derivative $d/d\tau$ to proper time. The constant terms
$T_{Mo}$ and $V_{Mo}$ have been absorbed on the right in energy
$E$. Then, starting from the Schwarzschild metric: in the plane of
the orbit, the metric in polar coordinates ($r$,$\phi$) simplifies
to
\begin{equation}
c^{2}d\tau^{2}=\alpha_{s}c^{2}dt^{2}-\frac{dr^{2}}{\alpha_{s}}-r^{2}d\phi^{2}.\label{eq:S-metric planar}
\end{equation}
Note that the \textquotedbl{}coordinate\textquotedbl{} length $r$
and time $t$ relate to the observer's reference frame at infinity.
The Schwarzschild dilation parameter is defined
\begin{equation}
\alpha_{s}(r)=1-\frac{r_{s}}{r},\label{eq:Schwarz alpha}
\end{equation}
where $r_{s}=2GM/c^{2}$ is the Schwarzschild radius. Taking in (\ref{eq:S-metric planar})
on both sides the derivative with respect to coordinate time $t$
and identifying the constant of motion $\dot{\tau}/\alpha_{s}=k$,
yields, after some manipulations \cite{Telkamp (2012)}, the \textquotedbl{}Schwarzschild
energy equation\textquotedbl{} for the orbit
\begin{equation}
\frac{{\scriptstyle \frac{1}{2}}m\dot{r}^{2}}{\alpha_{s}^{3}}+\frac{{\scriptstyle \frac{1}{2}}mr^{2}\dot{\phi}^{2}}{\alpha_{s}^{2}}-\frac{{\scriptstyle \frac{1}{2}}mc^{2}}{\alpha_{s}}={\scriptstyle -\frac{1}{2}}k^{2}mc^{2}=E.\label{eq:Schwarz energy eq}
\end{equation}
At this point, we introduce the \textit{relational dilation parameter},
the ratio of the observer's potential at position $r_{obs}$ and the
proper potential at the position $r$ of the test particle 
\begin{equation}
\alpha_{{\scriptscriptstyle R}}(r_{obs},r)\triangleq\frac{\varphi_{obs}(r_{obs})}{\varphi_{prop}(r)}.\label{eq:alpha rel}
\end{equation}
Using (\ref{eq:phio}), we specifically obtain for the Schwarzschild
case ($r_{obs}\rightarrow\infty$)
\begin{equation}
\hat{\alpha}_{{\scriptscriptstyle R}}(r)\triangleq\frac{\varphi_{obs}(\infty)}{\varphi_{prop}(r)}=\frac{\varphi_{o}}{\varphi_{o}+\varphi_{{\scriptscriptstyle M}}(r)}=\frac{1}{1+\frac{r_{s}}{r}}\approx1-\frac{r_{s}}{r}=\alpha_{{\scriptstyle {\scriptscriptstyle S}}}(r).\label{eq:alpha rel schw}
\end{equation}
$\alpha_{{\scriptstyle {\scriptscriptstyle S}}}(r)$ is virtually
identical to $\hat{\alpha}_{{\scriptscriptstyle R}}(r)$ for any admissible
value of $r$, since $r_{s}/r$ is generally extremely small. Note
that $\hat{\alpha}_{{\scriptscriptstyle R}}$ fits the Machian model,
since $m/\hat{\alpha}_{{\scriptscriptstyle R}}=m+\mu$. Replacing
$\alpha_{s}$ by $\hat{\alpha}_{{\scriptscriptstyle R}}$ converts
the Schwarzschild energy equation (\ref{eq:Schwarz energy eq}) into
the equivalent \textquotedbl{}relational energy equation\textquotedbl{}
for the Schwarzschild case
\begin{equation}
{\scriptstyle \frac{1}{2}}(m+\mu)\frac{\dot{r}^{2}}{\hat{\alpha}_{{\scriptstyle {\scriptscriptstyle R}}}^{2}}+{\scriptstyle \frac{1}{2}}m\frac{r^{2}\dot{\phi}^{2}}{\hat{\alpha}_{{\scriptstyle {\scriptscriptstyle R}}}^{2}}+(m+\mu)\varphi_{o}=E.\label{eq:Rel energy eq}
\end{equation}
This is nearly the above Machian equation (\ref{eq:Mach energy}),
except that the velocities in (\ref{eq:Rel energy eq}) are divided
by $\hat{\alpha}_{{\scriptstyle {\scriptscriptstyle R}}}$. This points
at an isotropic transform between the particle's proper coordinates
and the observer's coordinates, according to
\begin{equation}
\frac{ds}{dt}=\hat{\alpha}_{{\scriptscriptstyle R}}\,\frac{d\sigma}{d\tau}=\frac{\varphi_{obs}}{\varphi_{prop}}\,\frac{d\sigma}{d\tau},\label{eq:vel transf}
\end{equation}
where $d\sigma$ and $ds$ represent a displacement in arbitrary direction
in proper and observer coordinates, respectively. Eq. (\ref{eq:vel transf})
shows that the observed velocity of the particle varies along with
the observer's potential. For a comoving observer the transform is
unity. Thus the transform models the effects of remote observation
in curved spacetime, the part that is missing in a purely Machian
approach. What (\ref{eq:vel transf}) really reflects, though, is
the relational spacetime metric between two potentials at different
positions. Gravitational time dilation according to the Schwarzschild
metric (\ref{eq:S-metric planar}) obeys $d\tau^{2}=\hat{\alpha}_{{\scriptstyle {\scriptscriptstyle R}}}dt^{2}$.
Then, considering (\ref{eq:vel transf}), the spatial relational metric
must be $d\sigma^{2}=\hat{\alpha}_{{\scriptstyle {\scriptscriptstyle R}}}^{-1}ds^{2}$,
which implies isotropic length contraction. From this we derive the
\textit{isotropic relational metric} between two potentials $\varphi_{{\scriptscriptstyle A}}$
and $\varphi_{{\scriptscriptstyle B}}$ at arbitrary positions $A$
and $B$ 
\begin{equation}
\varphi_{{\scriptscriptstyle A}}\, d\sigma_{{\scriptscriptstyle B}}^{2}=\varphi_{{\scriptscriptstyle B}}\, d\sigma_{{\scriptscriptstyle A}}^{2},\label{eq:ds contr-1-1}
\end{equation}
\begin{equation}
\varphi_{{\scriptscriptstyle B}}\, d\tau_{{\scriptscriptstyle B}}^{2}=\varphi_{{\scriptscriptstyle A}}\, d\tau_{{\scriptscriptstyle A}}^{2}.\label{eq:dt dilation-1-1}
\end{equation}
This metric satisfies the relational principle: for a particle moving
from a fixed observer potential toward vacuum infinity ($\varphi_{prop}\rightarrow0$),
the proper units gradually dilute, meaning unbound increase of the
observed proper unit length and proper clock rate. 

The relational equation for the Schwarzschild case (\ref{eq:Rel energy eq})
can be generalized to arbitrary background potential $\varphi_{b}$
(by replacing $m$ by $\mu_{mb}=m\varphi_{b}/\varphi_{o}$) and arbitrary
observer potential $\varphi_{obs}$ (replacing $\hat{\alpha}_{{\scriptstyle {\scriptscriptstyle R}}}$
by $\alpha_{{\scriptstyle {\scriptscriptstyle R}}}$), yielding 
\begin{equation}
{\scriptstyle \frac{1}{2}}(\mu_{mb}+\mu)\frac{\dot{r}^{2}}{\alpha_{{\scriptstyle {\scriptscriptstyle R}}}^{2}}+{\scriptstyle \frac{1}{2}}\mu_{mb}\frac{r^{2}\dot{\phi}^{2}}{\alpha_{{\scriptstyle {\scriptscriptstyle R}}}^{2}}+(\mu_{mb}+\mu)\varphi_{o}=E.\label{eq:Rel energy eq-1}
\end{equation}

\noindent The question is how the relational transform of the Machian
equation extents to other spacetimes. Yet, the Schwarzschild example
shows that the generalized relational equation (\ref{eq:Rel energy eq-1})
covers both the applicable GR metric (subcase $\varphi_{b}=\varphi_{o}$)
and the applicable Machian equation (subcase $\varphi_{obs}=\varphi_{prop}$),
thus reconciling the two.

\end{document}